\documentclass[12pt,twoside]{article}
\usepackage{fleqn,espcrc1}

\usepackage{graphicx}
\usepackage[figuresright]{rotating}

\newcommand{\AmS}{{\protect\the\textfont2pine
  A\kern-.1667em\lower.5ex\hbox{M}\kern-.125emS}}

\title{Single target-spin asymmetries in semi-inclusive pion 
electroproduction on longitudinally polarized protons\thanks{talk 
presented by K.~Oganessyan at the Workshop on the structure ot the 
Nucleon (N$^{99}$), Frascati, June 7-9, 1999.}}

\author{A.M.~Kotzinian\address{CERN, CH-1211, Geneva 23, Switzerland}
\thanks{On leave of absence from Yerevan Physics Institute, Alikhanian 
Br.2, AM-375036 Yerevan, Armenia}\thanks{JINR, RU-141980 Dubna, Russia}, 
K.A.~Oganessyan\address{INFN-Laboratori Nazionali di Frascati, I-00040, 
Enrico Fermi 40, Frascati, Italy\\}$^{\dag}$, H.R.~Avakian$^{b \dag}$, 
E.~De~Sanctis$^{b}$}
       
\begin{document}

% typeset front matter
\maketitle

\begin{abstract}
\small{
We evaluate the single target-spin {\it sin}$\,\phi_h$ and {\it
sin}$\,2\phi_h$ azimuthal asymmetries in the semi-inclusive deep
inelastic lepton scattering off longitudinally polarized proton 
target under HERMES kinematic conditions. A good agreement with 
the HERMES data can be achieved using only the twist-2 distribution 
and fragmentation functions.\\ 
}
\end{abstract}

%\section{FORMAT}
Significant single-spin asymmetries have been observed
in experiments with transversely polarized proton and anti-proton
beams~\cite{E704}. 
Recently new experimental results on azimuthal
asymmetries became available. Specifically, the first
measurements of single target-spin azimuthal asymmetries of 
pion production in semi-inclusive deep inelastic scattering  
(SIDIS) of leptons off a longitudinally polarized target at 
HERMES~\cite{H99} and off a transversely polarized target at 
SMC~\cite{BRAV}, and the observation of the azimuthal correlations 
for particles produced from opposite jets in $Z$ decay at 
DELPHI~\cite{EFR}.  

In this note we present estimates of the 
single spin azimuthal asymmetry in the 
SIDIS on a longitudinally polarized nucleon target for the HERMES
kinematic conditions. Our approach is based on the parton model 
description of polarized SIDIS~\cite{AK}. The cross-section contains
the $(1/Q)^0$-order terms coming from leading dynamical twist-two 
distribution and fragmentation functions (DF's and FF's) as well as
$(1/Q)$-order kinematic twist-three terms arising due to the intrinsic
transverse momentum of the quark in the nucleon. We will neglect the
$(1/Q)$-order contributions of the higher twist DF's and FF's
obtained in~\cite{TM}. Thus, our approach is similar to that
of~\cite{CAHN} in describing the {\it cos}$\,\phi_h$ asymmetry
in unpolarized SIDIS.

\begin{figure}[ht]
\begin{center}
\includegraphics[width=8.cm]{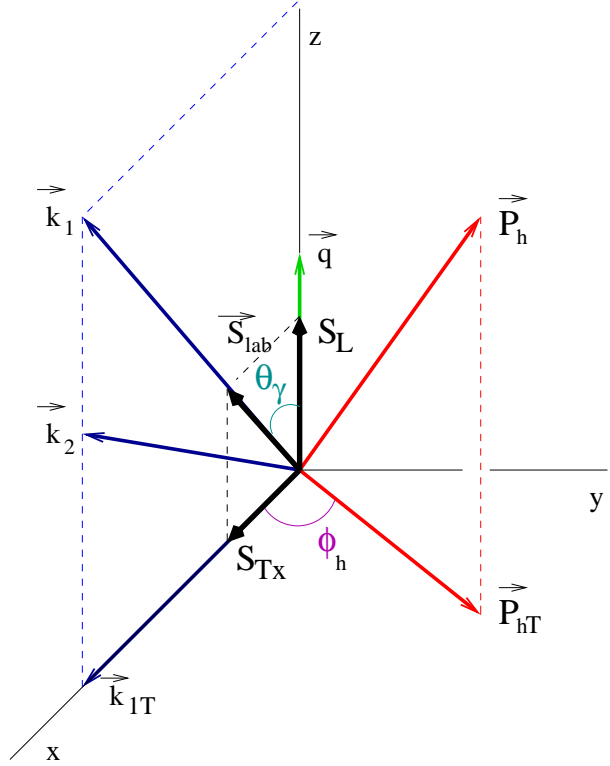}
\caption{The definition of the azimuthal angle $\phi_h$ and the target
  polarization components in virtual photon frame. }
\label{fb1}
\end{center}
\end{figure}

Let $k_1$ ($k_2$) be the initial (final) momentum of the incoming
(outgoing) charged lepton, $Q^2=-q^2$, $q=k_1-k_2$ -- the momentum of the
virtual photon, $P$ and $P_h$  ($M$ and $M_h$) -- 
the target and final hadron momentum (mass),
$x=q^2/2(Pq)$, $y=(Pq)/(Pk_1)$, $z=(PP_h)/(Pq)$,
$P_{hT}$ ($k_{1T}$) -- the hadron (lepton) transverse with respect to
virtual photon momentum direction and $\phi_h$ -- the azimuthal
angle between $P_{hT}$ and $k_{1T}$ around the virtual photon direction. 
Note that the azimuthal angle of the transverse (with respect to the 
virtual photon) component of the target polarization, $\phi_S$,
is equal to 0 ($\pi$) 
for the target polarized parallel (antiparallel) to the beam 
%\cite{OABK}  
 (Fig.~\ref{fb1}).
%\begin{figure}[ht]
%\vspace{-1.5 cm}

We use the approach developed in~\cite{KM} and consider the
cross-section integrated with different weights depending on the final
hadron transverse momenta $w_i(P_{hT})$
\footnote{More details can be found in~\cite{OABK}.}:  

\begin{equation}
\Sigma_i={ Q^2 y\over 2 \pi\alpha^2 } \int d^2P_{hT}\, w_i(P_{hT})\,d\sigma, 
\end{equation}
with $w_1(P_{hT}) = 1$, $w_2(P_{hT}) = \vert P_{hT} \vert\sin \phi_h/{M_h}$ 
and $w_3(P_{hT}) = {\vert P_{hT} \vert}^2 \sin 2\phi_h/{2MM_h}$.
Considering only the twist-two contributions, we have:  
\begin{equation}
\label{WI1}
\Sigma_1 = (1+(1-y)^2)\,f_1(x) D_1(z),  
%\nonumber \\
%&& \mbox{}
\end{equation}
where $f_1(x)$ and $D_1(z)$ are the usual unpolarized DF's and FF's. 
Moreover
\begin{equation}
\label{WI2}
\Sigma_2 = \Sigma_{2L}+\Sigma_{2T},
\end{equation}
where
\begin{equation}
\Sigma_{2L}=- 8S_L {M \over Q}\,(2-y)\sqrt{1-y} \, z
h^{\perp (1)}_{1L}(x) H_1^{\perp (1)}(z)   
\end{equation}
is the $(1/Q)$-order contribution from twist-two DF 
$h^{\perp(1)}_{1L}(x)$ and FF $H_1^{\perp (1)}(z)$ 
arising due to intrinsic transverse momentum and
\begin{equation}
\Sigma_{2T}= 2 S_{T\,x}\,(1-y)\,z
h_1(x) H_1^{\perp (1)}(z) 
\label{RRR}
\end{equation}
is arising due to the small ($\sim (1/Q)$) transverse component of 
the target 
polarization ($S_{T\,x}$) \cite{AK,OABK}. Finally 
% (see fig.~\ref{fb1}) .  
\begin{equation}
\label{WI3}
\Sigma_3 = 8 S_L (1-y) \, z^2   h_{1L}^{\perp(1)}(x)
H_1^{\perp (1)}(z).  
\nonumber \\  
\label{RRR1}
\end{equation} 
The weighted cross sections involve the $p_T^2$ ($k_T^2$) moment of 
the DF's (FF's), defined as
\begin{equation}
h_{1L}^{\perp (1)}(x) \equiv \int d^2p_T\,{\left(\frac{p_T^2}{2M^2}
\right)}\, h_{1L}^{\perp}(x, p_T^2), 
\label{WD} 
\end{equation}
\begin{equation}
H_1^{\perp (1)}(z) \equiv z^2 \int d^2k_T\,{\left(\frac{k_T^2}{2M^2_h}
\right)} H_1^{\perp}(z, z^2 k_T^2).  
\label{WF}
\end{equation}
We note that $h^{\perp}_{1L}(x)$ and $h_1(x)$ describe the
quark transverse spin distribution in the longitudinally
and transversely polarized nucleon respectively, while $H_1^{\perp}(z)$ 
describes 
the analyzing power of transversely polarized quark fragmentation
(Collins effect)~\cite{COL}. 

The single target-spin asymmetries for SIDIS on a longitudinally polarized 
target are defined as 
\begin{equation}
\langle \frac{\vert P_{hT}\vert}{M_h} \sin \phi_h \rangle \equiv 
\frac{\int d^2P_{hT} \frac{\vert P_{hT}\vert}{M_h}
\sin \phi_h \left(d\sigma^{+}-d\sigma^{-}\right)}
{\int d^2P_{hT} \left(d\sigma^{+} + d\sigma^{-}\right)},
\end{equation}
\begin{equation}
\langle \frac{{\vert P_{hT}\vert}^2}{MM_h} \sin 2\phi_h \rangle \equiv 
\frac{\int d^2P_{hT} \frac{{\vert P_{hT}\vert}^2}{MM_h}
\sin 2\phi_h \left(d\sigma^{+}-d\sigma^{-}\right)}
{\int d^2P_{hT} \left(d\sigma^{+} + d\sigma^{-}\right)},
\end{equation}
where $+ (-)$ denotes positive (negative) longitudinal polarization of the 
target.  
Using $\Sigma_{1,2,3}$ one can see that for both polarized and unpolarized
lepton these asymmetries are given by
\begin{equation}
\langle \frac{\vert P_{hT}\vert}{M_h} \sin \phi_h \rangle (x,y,z) = 
\frac{\Sigma_2(x,y,z)}{\Sigma_1(x,y,z)}
\label{AS}
\end{equation}
\begin{equation}
\langle \frac{{\vert P_{hT}\vert}^2}{MM_h} \sin 2\phi_h \rangle (x,y,z) = 
\frac{\Sigma_3(x,y,z)}{\Sigma_1(x,y,z)}. 
\label{AS1}
\end{equation}
We use the non-relativistic approximation  
$h_1(x) = g_1(x)$, the upper limit from Soffer's inequality
~\cite{SOF} $h_1(x) =  (f_1(x)+g_1(x))/2$,  and the relation
between $h_{1L}^{\perp(1)}(x)$ and $h_1(x)$ ~\cite{TM} obtained by
neglecting the interaction dependent twist-three part of the DF and
the term proportional to the current quark's mass: 
\begin{equation}
%h_L(x) = 2 x \int_{x}^1 dy {h_1(y) \over y^2} , \\    
%\label{HL} 
h_{1L}^{\perp (1)}(x) = - x^2 \int_{x}^1 dy {h_1(y) \over y^2}.   
\label{H1L}
\end{equation}
We took the parameterisations of DF's $f_1(x)$ and $g_1(x)$ 
from Ref.~\cite{BBS}. 
\begin{figure}[htb]
%\begin{minipage}[t]{80mm}
\begin{center}
\includegraphics[width=12.cm]{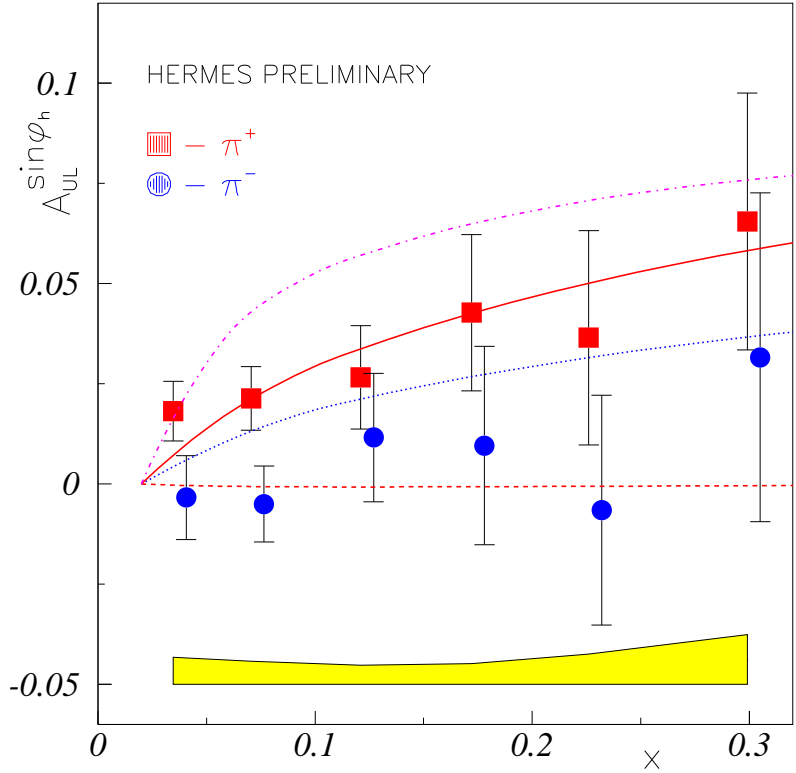}
\caption{The $A^{\sin\phi_h}_{UL}(x)$ asymmetry of $\pi^{\pm}$ production. 
The continuous ($\pi^{+}$) and dashed ($\pi^{-}$) curves correspond to  
$M_C=0.7$ GeV, $h_1=g_1$; dotted ($\pi^{+}$) and dot-dashed ($\pi^{+}$) 
to $M_C=0.3$ GeV, $h_1=g_1$ and $M_C=0.7$GeV $h_1=(f_1+g_1)/2$, respectively. 
} 
\label{fig:f1}
%\end{minipage}
\end{center}
\end{figure}
%\vspace{-1.5 cm}
To calculate the T-odd FF $H_1^{\perp (1)}(z)$ we adopt the Collins
parameterisation~\cite{COL} for the analyzing power of 
transversely polarized quark fragmentation
\begin{equation}
A_C(z,k_T) \equiv \frac{\vert k_T \vert}{M_h}\frac{H_1^{\perp}(z,k_T^2)}
{D_1(z,k_T^2)} = \frac{M_C\,\vert k_T \vert}{M_C^2+k_T^2}
\label{H1T}
\end{equation}
and assume a Gaussian parameterisation of the unpolarized FF~\cite{KM} 
with $\langle z^2 k_T^2 \rangle = b^2 $ 
(in the numerical calculations we use 
$b = 0.5$ GeV \cite{E665}). For $D_1^{\pi^{\pm}} (z)$ we use the 
parameterisation from Ref.~\cite{REYA}. 
%In Eq.(\ref{H1T}) $M_C \simeq 0.3\div 1.0$~GeV is a typical hadronic mass. 

The $A^{\sin\phi_h}_{UL}(x)$ asymmetry for $\pi^{\pm}$
production on the proton target is obtained from the 
defined asymmetry (Eq.(\ref{AS})) by the relation $A^{\sin\phi_h}_{UL} 
\approx {2M_h \over {\langle P_{hT}\rangle }} \langle \frac{\vert 
P_{hT}\vert}{M_h} \sin \phi_h \rangle$ and is presented in Fig.~\ref{fig:f1} 
in comparison with preliminary HERMES data~\cite{H99}. 
The data corresponds 
to $Q^2 \geq 1$ GeV$^2$, $E_{\pi} \geq 4$ GeV, and the ranges 
$0.2 \leq z \leq 0.7$, $0.2 \leq y \leq 0.8$. The theoretical curves 
are calculated by integrating over the same ranges with 
$\langle P_{hT} \rangle=0.52$ GeV, $\langle P_{hT}^2 \rangle=0.35$ GeV$^2$. 
These average values of $P_{hT}$, $P_{hT}^2$  are obtained in mentioned 
kinematics assuming a Gaussian parameterisation of DF's and FF's with 
$a=0.7$ GeV ($\langle p_T^2 \rangle=a^2 $)~\cite{E665}.   
From Fig.~\ref{fig:f1} one can see that a good agreement with HERMES
data~\cite{H99} can be achieved by varying $h_1(x)$ and $M_C$. 
Note that the main effect comes from the $\Sigma_{2L}$ term, the 
contribution of $\Sigma_{2T}$ is about $20 \div 25 \%$. 
 \begin{figure}[htb]
%\begin{minipage}[t]{75mm}
\begin{center}
\includegraphics[width=10.cm]{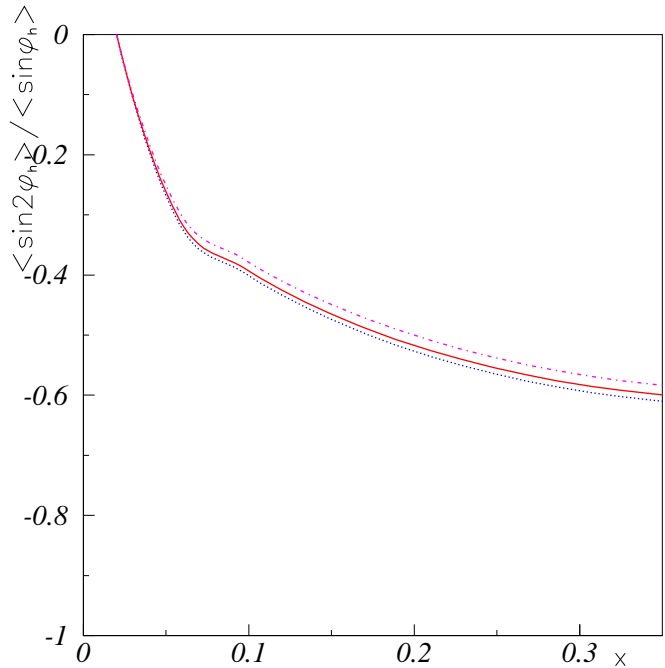}
\caption{
The ratio of the amplitudes of the $\sin2\phi_h$ and $\sin\phi_h$ 
single target-spin asymmetries for $\pi^{+}$ production. 
The curves have the same notations as in the Fig.~\ref{fig:f1}. 
}
\label{fig:f2}
%\end{minipage}
\end{center}
\end{figure}
%\vspace{-1.5 cm}

We calculate the $sin2\phi_h$-weighted asymmetry in the same
manner as well and show that the amplitude of the $sin2\phi_h$ 
modulation is about a factor of 2-3 smaller than that of the $sin\phi_h$ 
modulation (see Fig.~\ref{fig:f2}) in the HERMES kinematics. Note that the 
ratio of these asymmetries is almost independent of the choice of  
$h_1(x)$ and $M_C$.  
 
%\section{Conclusions}

In conclusion, the {\it sin}$\,\phi_h$ and {\it sin}$\,2\phi_h$
single target-spin asymmetries of SIDIS off longitudinally polarized
protons related to the time reversal odd FF was investigated. 
It was shown that the main $(1/Q)$-order contribution to the spin 
asymmetry arises from 
intrinsic $k_T$ effects similar to the $cos\phi_h$ asymmetry in 
unpolarized SIDIS. 
A good agreement with the HERMES data can be achieved using only the 
twist-2 DF's and FF's. The $(1/Q)^0$-order $sin 2\phi_h$ asymmetry, in
contrast to the naive expectations, is suppressed comparing
to the $(1/Q)$-order $sin \phi_h$ asymmetry at HERMES kinematics.   

%\section{Acknowledgements}
The authors would like to thank D.~Boer, R.~Jakob, and P.~Mulders for
useful discussions. 
The work of (K.O) and (H.A) was in part supported by the INTAS
contributions (contract number 93-1827) from the European Union. 

%\section*{References}

\end{document}